\documentclass[aps, twocolumn, a4paper, superscriptaddress]{revtex4}

\usepackage{graphicx}
\usepackage{float}
\usepackage{amsmath, amssymb}
\usepackage{lmodern}
\usepackage[T1]{fontenc}
\usepackage[utf8]{inputenc}
\usepackage[english]{babel}
\usepackage{color}

\def\lsim{\mathrel{\rlap{\lower4pt\hbox{\hskip1pt$\sim$}}
    \raise1pt\hbox{$<$}}}                
\def\gsim{\mathrel{\rlap{\lower4pt\hbox{\hskip1pt$\sim$}}
    \raise1pt\hbox{$>$}}}   

\begin{document}

\title{Rock-paper-scissors models with a preferred mobility direction}

\author{P.P. Avelino}
\affiliation{Departamento de F\'{\i}sica e Astronomia, Faculdade de Ci\^encias, Universidade do Porto, Rua do Campo Alegre s/n, 4169-007 Porto, Portugal}

\affiliation{Instituto de Astrof\'{\i}sica e Ci\^encias do Espa{\c c}o, Universidade do Porto, CAUP, Rua das Estrelas, 4150-762 Porto, Portugal}

\affiliation{School of Physics and Astronomy, University of Birmingham, Birmingham B15 2TT, United Kingdom}

\author{B.F. de Oliveira}
\affiliation{Departamento de Física, Universidade Estadual de Maringá, Av. Colombo 5790, 87020-900 Maringá, PR, Brazil}

\author{J.V.O. Silva}
\affiliation{Departamento de Física, Universidade Estadual de Maringá, Av. Colombo 5790, 87020-900 Maringá, PR, Brazil}

\begin{abstract}
We investigate a modified spatial stochastic Lotka-Volterra formulation of the rock-paper-scissors model using off-lattice stochastic simulations. In this model one of the species moves preferentially in a specific direction --- the level of preference being controlled by a noise strength parameter $\eta \in [0,1]$ ($\eta=0$ and $\eta=1$ corresponding to total preference and no preference, respectively)  --- while the other two species have no preferred direction of motion. We study the behaviour of the system starting from random initial conditions, showing that the species with asymmetric mobility has always an advantage over its predator. We also determine the optimal value of the noise strength parameter which gives the maximum advantage to that species. Finally, we find that the critical number of individuals, below which the probability of extinction becomes significant, decreases as the noise level increases, thus showing that the addition of a preferred mobility direction studied in the present paper does not favour coexistence.
\end{abstract}

\maketitle

\section{INTRODUCTION}

Cyclic predator-prey models, also known as rock-paper-scissors (RPS) models \cite{2002-Kerr-N-418-171,2007-Reichenbach-N-488-1046}, have provided insight into the role of non-hierarchical competition interactions in the preservation of coexistence, successfully reproducing some of the main properties of simple biological systems with cyclic selection \cite{1996-Sinervo-Nature-380-240,2002-Kerr-N-418-171,bacteria}. The simplest models in this class describe the dynamics of a three species population subject to cyclic interspecific competition (see also \cite{2008-Szabo-PRE-77-041919, 2010-Shi-PRE-81-030901, 2011-Allesina-PNAS-108-5638, 2012-Avelino-PRE-86-031119, 2012-Avelino-PRE-86-036112,  2012-Li-PA-391-125, 2012-Roman-JSMTE-2012-p07014, 2013-Lutz-JTB-317-286, 2013-Roman-PRE-87-032148, 2014-Cheng-SR-4-7486, 2016-Kang-Entropy-18-284, 2016-Roman-JTB-403-10, 2017-Brown-PRE-96-012147, 2017-Park-SR-7-7465, 2017-Bazeia-EPL-119-58003, 2017-Souza-Filho-PRE-95-062411, 2018-Avelino-PRE-97-032415, 2018-Shadisadt-PRE-98-062105, 2019-Park-Chaos-29-051105} for generalizations of the standard RPS model involving additional species and interactions and \cite{2014-Szolnoki-JRSI-11-0735, 2018-Dobramysl-JPA-51-063001} for recent reviews).

The role of mobility has been the subject of many studies which have shown that it may promote or jeopardize biodiversity (see, e.g., \cite{2007-Reichenbach-N-488-1046}). Although most of these studies considered a uniform isotropic mobility, it has been shown in \cite{2018-Avelino-PRE-97-032415} that a non-uniform anisotropic mobility may affect coexistence in a (four state) May-Leonard formulation of the RPS model using lattice based simulations. In this model the direction of motion for each individual was assumed to be the one with a larger density of preys in the surrounding neighborhood. In this context, anisotropic mobility has been shown to have a profound impact on the dynamics of the population and on the emerging spatial patterns.

In the present paper rather than attempting to simulate the ability of individuals to detect surrounding prey using their senses, as done in \cite{2018-Avelino-PRE-97-032415}, we investigate the potential impact of strong correlations between the motion of individuals in a given (large) neighbourhood. This is generally observed in nature, specially among the most developed species, as a result of predator-prey interactions. To this end, we shall investigate modified spatial stochastic RPS models in which individuals of one of the species move preferentially in a specific direction while the other two species have an isotropic mobility. In the present paper we shall consider a (three state) Lotka-Volterra formulation of the RPS model and perform off-lattice stochastic simulations, which, unlike lattice based ones, allow individuals to move in a continuous spatial area (see \cite{1995-Vicsek-PRL-75-1226, 2010-Ni-C-20-045116, 2010-Ni-PRE-82-066211, 2017-You-JTB-435-78, 2018-Avelino-EPL-121-48003} for numerical studies using simulations of this type).

The outline of this paper is as follows. In Sec. \ref{model} we start by describing the RPS models with a preferred mobility direction that shall be investigated here. In Sec. \ref{results} we present the results, characterizing the main features associated to the time and spatial dynamics of the populations and showing how the existence of a preferred mobility direction of one species investigated in the present paper can both have a positive impact on that species abundance and a negative impact on the preservation of coexistence. Finally, we conclude in Sec. \ref{conclusion}.

\section{THE MODEL \label{model}}

Here, we consider a modified spatial stochastic Lotka-Volterra formulation of the RPS model. To this end, we shall perform off-lattice simulations in which the individuals of the various species (labelled by the letters $A$, $B$ and $C$) are initially randomly distributed in a square-shaped cell of linear size, $L$, with periodic boundary conditions. At the start of the simulations the number of individuals of any of the three different species is the same ($N_A=N_B=N_C=N/3$, where $N$ represents the total number of individuals). The fractional abundance of individuals shall be defined by $\rho_i=N_i/N$, with $i \in {A,B,C}$.

\begin{figure}[t]
	\centering
		\includegraphics{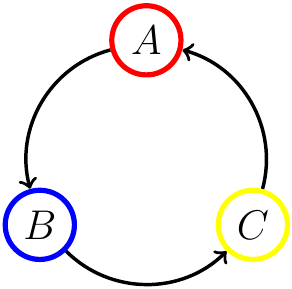} 
	\caption{Illustration of non-hierarchical predator-prey interactions in RPS models.}
	\label{fig1}
\end{figure}

At each simulation time step, a single individual (active) is chosen at random and one action --- either mobility or predation --- is selected, with any of these two possible actions carrying the same probability. Figure \ref{fig1} shows the standard scheme of cyclic predation employed in our model: $A$ predates $B$, $B$ predates $C$, and $C$ predates $A$.  Whenever predation is selected a circle with radius $\ell$ is drawn around the active individual and the nearest prey, if it exists, is replaced by an individual of the same species of the active individual --- if there is no prey inside the circle then nothing happens (note that in a Lotka-Volterra formulation of the RPS model, predation and reproduction take place simultaneously). On the other hand, if mobility is selected and the active individual belongs to species $B$ (blue) or $C$ (yellow), then it moves in a random direction with a step size $\ell = 2 \times 10^{-2}$. If the active individual belongs to species $A$ (red) then we shall consider two possible mobility implementations, which shall be referred as model I and model II. In model I the random direction associated to the mobility of species $A$ is restricted to an angle $\Delta \theta = \xi(t) \times \eta$ around the $\hat{x}$ direction, where $\xi$ is a random variable uniformly distributed on $[-\pi: \pi)$ and $\eta \in [0,1]$ is the noise strength. Note that for $\eta =1$ there is no preferred direction of motion while for $\eta = 0$ the individuals of the species $A$ move always in the same $\hat{x}$ direction. Model II, inspired in Vicsek's model \cite{1995-Vicsek-PRL-75-1226}, is similar to model I, except that the $\hat{x}$ direction is replaced by the average direction of motion of the individuals in a neighborhood of radius $r$ of the selected individual of the species $A$ --- for the sake of definiteness we shall consider $r=0.2$ throughout the paper, except if stated otherwise. One generation is defined as the time necessary for $N$ actions to take place.  

Except for the special case discussed in Figure \ref{fig3}, all simulations performed in the present paper consider random initial conditions and have a total time span of $1.1 \times 10^4$ generations --- the first $10^3$ generations being discarded in the derivation of our main results. The total number of individuals considered in the simulations is $N = 3 \times 10^4$, except in the study of extinction probability where different values in the interval $[99, 9999]$ have been employed.

\section{RESULTS \label{results}}

\begin{figure}[t]
	\centering
		\includegraphics[width=8.5cm]{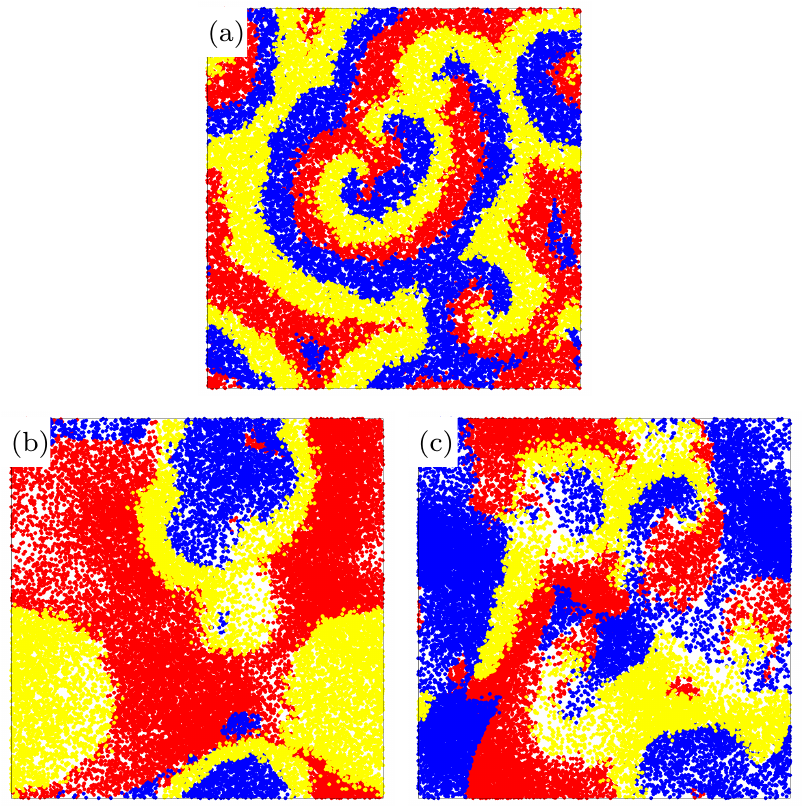} 
	\caption{Snapshots of spatial stochastic Lotka-Volterra numerical off-lattice simulations of RPS models with a preferred mobility direction. The snapshots were taken after $5 \times 10^3$ generations of simulations with $3 \times 10^4$ individuals and random initial conditions, considering: (a) the RPS model with isotropic mobility --- $\eta = 1.0$ (b) model I with $\eta=0.1$ (c) model II with $\eta = 0.1$.}
	\label{fig2}
\end{figure}

Figure \ref{fig2} presents three off-lattice simulation snapshots taken after $5 \times 10^3$ generations, considering: (a) the RPS model with isotropic mobility ($\eta = 1.0$)  \cite{2018-Avelino-EPL-121-48003} (b) model I with $\eta=0.1$ (c) model II with $\eta = 0.1$. Note that the distinct spiral patterns present in the top panel of Fig. \ref{fig2} (case (a)) are absent in the two bottom panels (cases (b) and (c)). On the other hand, the characteristic size of the spatial structures seems to decrease with $\eta$, being larger in cases (b) and (c) than in case (a). Also, in cases (b) and (c) there are regions with a large density of empty patches that do not seem to occur in case (a).

\begin{figure}[t]
	\centering
		\includegraphics[width=8.5cm]{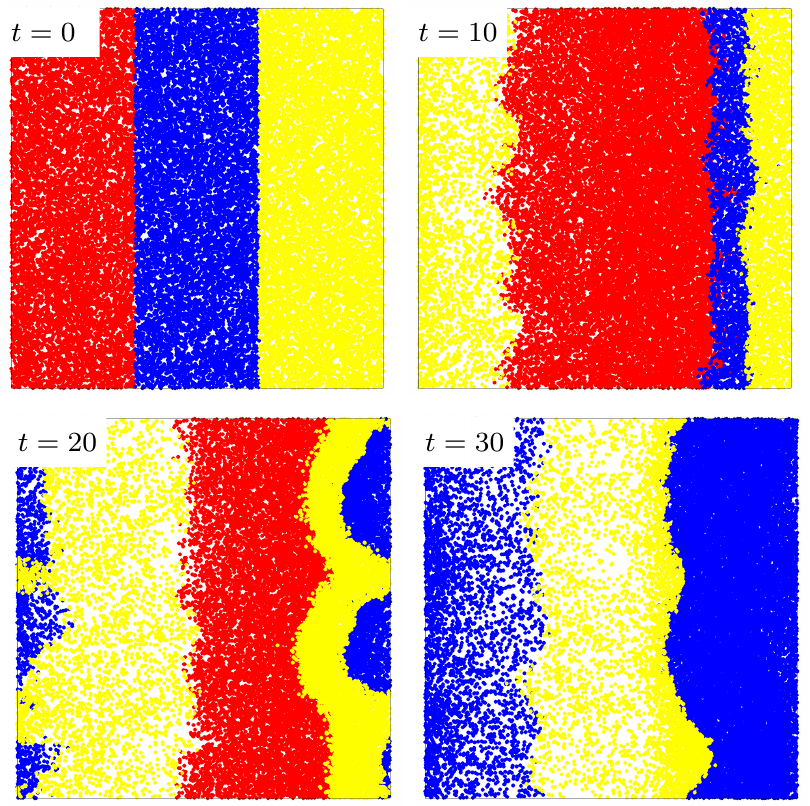} 
	\caption{Snapshots of the evolution of a single Lotka-Volterra off-lattice simulation of a RPS model with a preferred mobility direction (model I, with $\eta=0.1$ and $N=3 \times 10^4$), considering the initial configuration shown in the top left panel. The subsequent snapshots were taken after 10, 20, and 30 generations (top right, bottom left and bottom right panels, respectively). After 30 generations the red species is extinguished.}
	\label{fig3}
\end{figure}

In order to get a better understanding of the process responsible for the formation of regions with a high density of empty patches we consider a simulation of model I with $\eta=0.1$ in which individuals from the three species were initially distributed along three vertical strips (red, blue, yellow, respectively) as shown in Fig. \ref{fig3}. The preferred mobility of the red species in the $\hat{x}$ direction is responsible for an initial fast decrease of the blue species population and for the high density of empty patches in the boundary region separating the yellow and red species populations. When the blue region gets sufficiently  thin it becomes permeable to the passage of individuals of the yellow species which end up engulfing the individuals of the red species until it finally becomes extinct.

\begin{figure}[ht]
	\centering
	\includegraphics{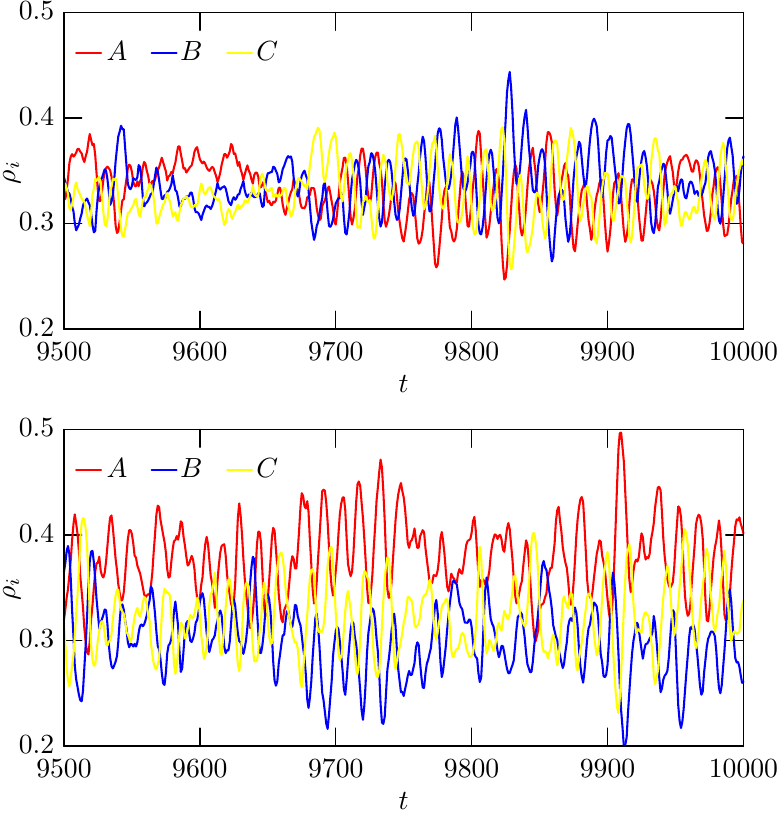}
	\caption{Evolution of the fractional abundances $\rho_i$ of the various species as a function of time for a single realization of our model considering $\eta= 1.0$ (top panel) and $\eta= 0.7$ (bottom panel), and random initial conditions.}
	\label{fig4}
\end{figure}

The evolution of the fractional abundances $\rho_i$ of the various species as a function of time is displayed in Figure \ref{fig4} for a single realization of model I considering $\eta= 1$ (top panel) and $\eta= 0.7$ (bottom panel),  and random initial conditions. Figure \ref{fig4} shows that for $\eta=1$ (top panel) all the abundances oscillate around the common average value of $1/3$, while for $\eta=0.7$ there is a clear advantage of the species $A$, both over its prey and its predator species. Moreover, the characteristic time and amplitude of the oscillations is larger in the case with $\eta=0.7$ than in the case with $\eta=1$. This is associated to the larger characteristic size of the spatial structures in the former case. The oscillations are also less sinusoidal for $\eta=0.7$ compared with the case with $\eta=1$.

\begin{figure}[ht]
	\centering
		\includegraphics{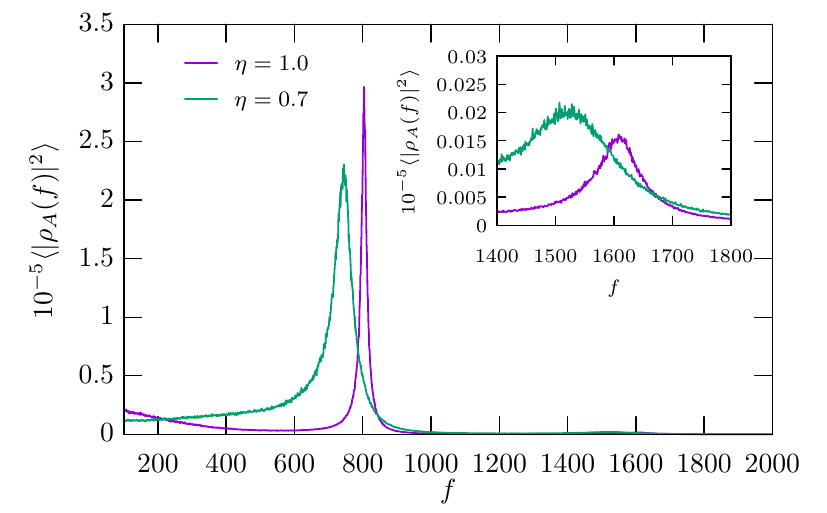}
	\caption{Power spectrum $\langle |\rho_A(f)|^2 \rangle$ for the time evolution of the fractional abundance $\rho_A$ of the red species for the cases with $\eta=0.7$ (solid green line) and $\eta=1$ (solid magenta line). The inset highlights the power around the second harmonic.}
	\label{fig5}
\end{figure}

The dependence of the characteristic frequency on $\eta$ may be further quantified using the power spectrum. The temporal discrete Fourier transform is defined as
\begin{equation}
	\rho(f) = \displaystyle\sum_{t=0}^{N_G-1}
\rho(t)e^{-2\pi i t f/N_G} \ ,
	\label{eq1}
\end{equation}
where $\rho(t)$ is the fractional abundance of a species, $N_G= 10^4$
generations and $f$ is the frequency.
Figure \ref{fig5} displays the power spectrum $\langle |\rho_A(f)|^2 \rangle$ for the time evolution of the fractional abundance $\rho_A$ of the red species for the cases with $\eta=0.7$ (solid green line) and $\eta=1$ (solid magenta line). It shows that maximum of the power occurs at a smaller frequency --- referred to as the fundamental frequency or first harmonic --- in the presence of a preferred directional mobility for species $A$ ($\eta=0.7$) than in the case where the mobility of all species is isotropic ($\eta=1$). On the other hand, the width of the power spectrum is larger in the former than in the later case. These properties of the power spectrum are a result of the larger characteristic time, size and amplitude of the oscillations observed for $\eta=0.7$. The inset in Figure \ref{fig5} highlights the power around the second harmonic. Note that the peak ratio between the first and second harmonics is larger for $\eta=1$ than for $\eta=0.7$, which is associated to the less sinusoidal nature of the later case compared to the former one (see \cite{2015A&A...576A..15R,2017A&A...599A...1S} for an application of the peak ratios in an astrophysical context).

\begin{figure}[t]
	\centering
		\includegraphics{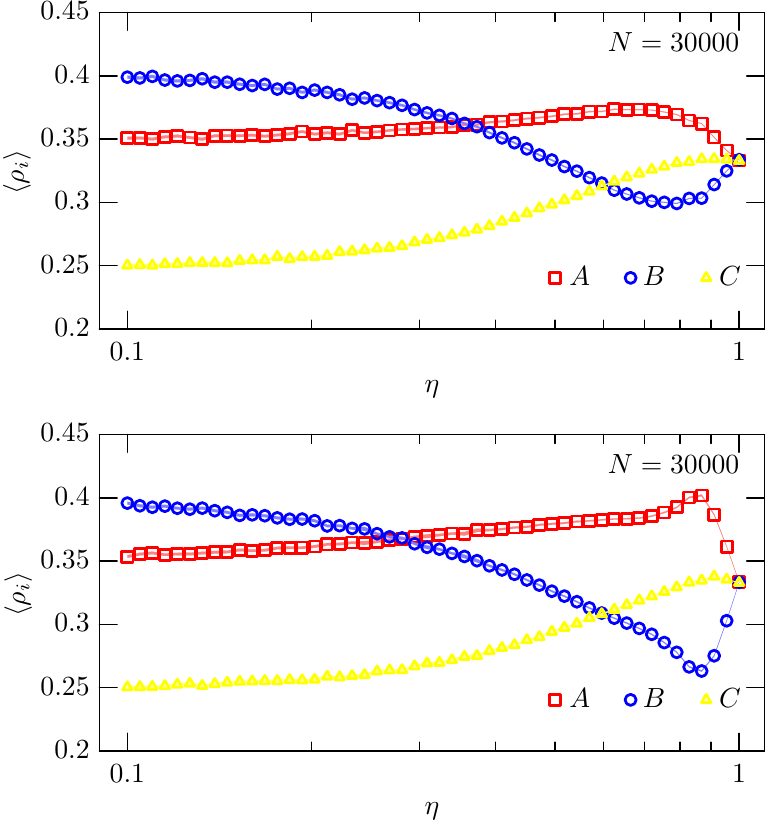}
	\caption{The average fractional abundances $\langle \rho_i \rangle$ of the species $i=A$, $B$ and $C$ obtained for model I (top panel) and model II (bottom panel) as a function of the noise parameter $\eta$. All simulations have a total number of individuals $N$ equal to $3\times 10^4$ and the average was performed over the last $5 \times 10^4$ generations of simulations with a time span equal to $5.1 \times 10^4$ generations. Notice that in both cases for $\eta < 1$ the species with an anisotropic mobility (species $A$) always has an advantage over its predator (species $C$).}
	\label{fig6}
\end{figure}

We also consider the dependency of the average fractional abundances $\langle \rho_i \rangle$ of the various species on the noise parameter $\eta$. To this end, we perform simulations with a total number of individuals $N$ equal to $3\times 10^4$ --- the average abundances are computed considering a time average over the last $5 \times 10^4$ generations of simulations with a time span equal to $5.1 \times 10^4$ generations. Figure \ref{fig6} shows the average fractional abundances $\langle \rho_i \rangle$ of the species $i=A$, $B$ and $C$ obtained for model I (top panel) and model II (bottom panel) as a function of the noise parameter $\eta$. Although for maximum noise strength ($\eta = 1$) all species have the same average fractional abundance, for $\eta < 1$ the species with an anisotropic mobility (species $A$) always has an advantage over its predator (species $C$) in both model I and model II (in model I the maximum advantage of the red species occurs for a noise strength $\eta \sim 0.7$, while for model II it occurs for a value of $\eta$ closer to unity). The results for model I and model II are qualitatively similar, except for values of $\eta$ close to unity. We also verified that for $r \gsim 0.5$ models I and II produce similar results.

For large values of the noise strength ($0.35 \lsim \eta \lsim 1$) the red species is the dominant one, surpassing both its prey and its predator. For small values of the noise strength ($\eta \lsim 0.35$) the preying efficiency of the red species is negatively affected by the anisotropic mobility and the blue species becomes the most abundant (at the expense of the yellow species). Species $C$ and species $B$ are the least abundant for $\eta \lsim 0.6$ and $0.6 \lsim \eta \lsim 1$, respectively.

\begin{figure}[t]
	\centering
		\includegraphics{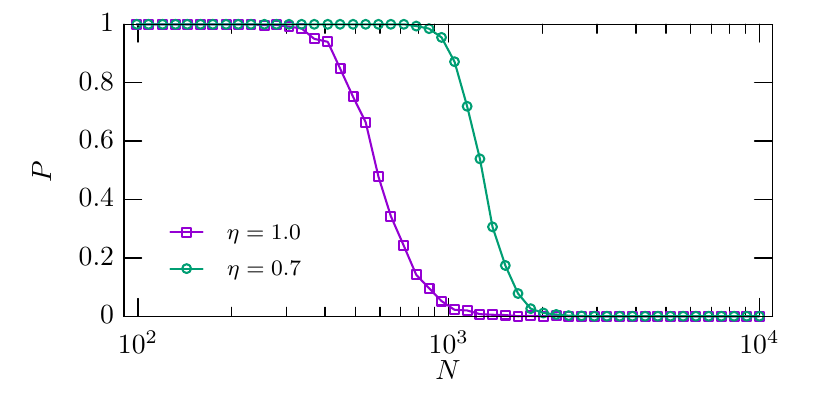}
	\caption{The extinction probability $P$ is depicted as a function of the total 
number of individuals $N$. Each point was obtained from $10^3$ simulations. Notice, that the critical number of individuals, below which the probability of extinction becomes significant, decreases as $\eta$ increases. This result implies that the anisotropic mobility considered in the present paper does not favour coexistence.}
	\label{fig7}
\end{figure}

The impact of the existence of a preferred mobility direction (for species $A$) on coexistence may be studied by estimating the extinction probability $P$ as a function of the number of individuals $N$. This is shown in Fig. \ref{fig7}. Each point was obtained from $10^3$ simulations and, consequently, the one-sigma uncertainty in the value of $P$ may be estimated as $[P(1-P)/10
^3]^{1/2}$, with a maximum of approximately $0.016$ for $P=0.5$. Figure \ref{fig7} shows that the critical number of individuals below which the probability of extinction becomes significant decreases as the level of anisotropy decreases (or, equivalently, as $\eta$ increases) --- note that the results obtained for $\eta = 1.0$ agree well with those presented in \cite{2018-Avelino-EPL-121-48003}. However, for $\eta= 0.7$ a greater number of individuals is necessary in order for extinction to be avoided. In fact, we have found that the critical number of individuals $N_c$ above which the probability of extinction becomes significant ($P(N_c) \equiv 0.5$) is approximately equal to $N_c = 9.4 \times 10^2$ for $\eta = 1.0$, and $N_c = 1.2 \times 10^3$ for $\eta = 0.7$. This behaviour implies that the preferred mobility direction of species $A$ considered in the present paper does not favour coexistence.

\section{CONCLUSIONS \label{conclusion}}

In this work we investigated the dynamics of a population of individuals from three different species in the context of a spatial stochastic Lotka-Volterra formulation of the RPS model where one of the species has a preferred mobility direction. This has been accomplished using off-lattice stochastic simulations, starting from random initial conditions. We have shown that the anisotropic mobility has a significant impact on the spatial patterns which form as a result of the population dynamics, with the distinct spiral patterns, common in the isotropic case, becoming unrecognizable in the present of a significant asymmetric mobility. We characterized the relative abundance of the three species as a function of the noise level, showing, in particular, that the species with asymmetric mobility has always an advantage over its predator. We have also determined the optimal value of the noise strength parameter which is associated to the maximum advantage of that species relative to the other two. Finally, we have found that the threshold number of individuals, below which the probability of extinction becomes significant, decreases as the noise level increases, thus showing that the preferred mobility direction studied in the present paper does not favour coexistence.

\section*{ACKNOWLEDGMENTS}

\begin{acknowledgments}
P.P.A. acknowledges the support from FCT through the Sabbatical Grant No. SFRH/BSAB/150322/2019 and through the Research Grants No. UID/FIS/04434/2019, UIDB/04434/2020 and UIDP/04434/2020. B.F.O. and J.V.O.S. thank CAPES - Finance Code 001, Funda\c c\~ao Arauc\'aria, and INCT-FCx (CNPq/FAPESP) for financial and computational support.
\end{acknowledgments}

\end{document}